\documentclass{ifacconf}

\usepackage{graphicx}
\usepackage[dvips]{epsfig}

\usepackage{amssymb,amsmath,amsfonts,subfigure,enumerate}
\usepackage{accents}
\usepackage{multirow}
\usepackage{dsfont}
\usepackage{natbib}
\usepackage{color}

\def\la{\lambda}

\def\ve{\varepsilon}

\def\re{\mathop{Re}\nolimits}

\def\om{\omega}

\def\g{\mathcal G}

\def\r{\mathbb R}

\def\be{\beta}

\def\S{\mathfrak S}

\newcommand{\dfb}{\stackrel{\Delta}{=}}

\def\E{\mathcal E}
\def\V{\mathcal V}

\def\epf{$\blacksquare$}

\def\be{\begin{equation}}
\def\ee{\end{equation}}
\def\ben{\begin{equation*}}
\def\een{\end{equation*}}

\begin{document}

\begin{frontmatter}
\title{Simple synchronization protocols for heterogeneous networks: beyond passivity (extended version)}

\thanks[footnoteinfo]{
Partial funding was provided by STW (project 13712).
E-mails: {\tt\small anton.p.1982@ieee.org, m.mazo@tudelft.nl}
}

\author[DCSC,IPME]{Anton V. Proskurnikov}
\author[DCSC]{Manuel Mazo Jr}

\address[DCSC]{Delft Center for Systems and Control (DCSC), Delft University of Technology, The Netherlands}
\address[IPME]{ITMO University \& Institute for Problems of Mechanical Engineering (IPME RAS), St. Petersburg, Russia}

\begin{abstract}
Synchronization among autonomous agents via local interactions is one of the benchmark problems in multi-agent control. Whereas
synchronization algorithms for identical agents have been thoroughly studied, synchronization of heterogeneous networks still remains a challenging problem.
The existing algorithms primarily use the internal model principle, assigning to each agent a local copy of some dynamical system (internal model). Synchronization
of heterogeneous agents thus reduces to global synchronization of identical generators and local synchronization between the agents and their internal models.
The internal model approach imposes a number of restrictions and leads to sophisticated dynamical (and, in general, nonlinear) controllers.
At the same time, passive heterogeneous agents can be synchronized by a very simple linear protocol, which is used for consensus of first-order integrators.
A natural question arises whether analogous algorithms are applicable to synchronization of agents that do not satisfy the passivity condition.
In this paper, we study the synchronization problem for heterogeneous agents that are not passive but satisfy a weaker input feedforward passivity (IFP) condition.
We show that such agents can also be synchronized by a simple linear protocol, provided that the interaction graph is strongly connected and the couplings are sufficiently weak.
We demonstrate how stability of cooperative adaptive cruise control algorithms and some microscopic traffic flow models reduce to synchronization of heterogeneous IFP agents.
\end{abstract}

\end{frontmatter}


\section{Introduction}

As the influential monograph~\citep{StrogatzSync} states, ``the tendency to synchronize is one of the most pervasive drives in the universe, extending from atoms to animals''. Synchrony among subsystems (agents, cells) of a complex system is a basic principle, which explains many natural phenomena~\citep{StrogatzSync} and has found numerous applications in engineering~\citep{MesbahiEgerBook,Murray:07,Wu2007,RenBeardBook}. Establishing synchronization (consensus) is considered now as a benchmark problem in multi-agent control and has been thoroughly examined in the recent decades.

Most of the attention has been paid to synchronization among \emph{identical} agents. The protocols establishing synchronization among single integrators are usually
based on the idea of contraction: the convex hull, spanned by the agents' states, is shrinking until it collapses into a singleton~\citep{Muenz:11}. An alternative approach is based on
convergence criteria for infinite matrix products~\citep{RenBeardBook}. The protocols for synchronization of agents, obeying higher order equations, are similar in spirit to first-order algorithms. Synchronization of linear and linearly coupled agents is often analyzed via the spectral decomposition of the Laplacian matrix~\citep{Murray:07,LiDuanChen:10,RenBeardBook,RenCaoBook}.
Nonlinear protocols are usually examined by Lyapunov methods~\citep{RenCaoBook},
employing, among others, the Kalman-Yakubovich-Popov lemma~\citep{FanHarryJacquelien:14,ProMatv:15}, contraction theory~\citep{Bernardo:11} and the idea of incremental passivity~\citep{StanSepulchre:07,ProZhangCao:2015,LiuHillZhao:15}.

However, in practice autonomous agents are usually {heterogeneous}. Algorithms for output synchronization of non-identical agents have been proposed quite recently and most of them employ the \emph{internal model principle}~\citep{WielandAllgo,DePersisBayu:14,IsidoriMarconi:2014,BidramLewis:14,LiuDePersisCao:15}, assigning to each agent a virtual copy of some dynamical system, referred to as
the \emph{internal model} or the \emph{local reference generator}. The control algorithm then consists of two layers: a protocol, synchronizing the (identical) reference generators and local \emph{model-matching} controllers, synchronizing the agents to their generators.

The general internal model approach has, however, several disadvantages. Being formally decentralized, its implementation assumes that the agents share the same internal model and are able to
match it (e.g. in the case of linear agents the Francis regulator equations should be solvable~\citep{WielandAllgo,LiuDePersisCao:15}). Hence design of an algorithm requires to know the global information about the network. Unlike many synchronization algorithms for identical agents~\citep{Murray:07,LiDuanChen:10,RenBeardBook,RenCaoBook} that use only \emph{relative measurements}, that is, the deviations between an agent's output and the outputs of its neighbors, the model-matching controllers need access to the absolute outputs of the agents.
Dealing with mobile robots, this implies that agents have to measure their positions and/or velocities in the \emph{global} frame of reference.

At the same time, synchronization among heterogeneous \emph{passive} agents (e.g. mechanical systems in the Euler-Lagrange form) can be established by the same
simplest protocols~\citep{Pogromsky:01,Arcak:07,ChopraSpongBook} as used to synchronize single integrator agents~\citep{Murray:07}.
Such a protocol does not require any knowledge of the agents' dynamics (except for their passivity) and uses only deviations between the agents' outputs, but not the outputs themselves.

Thus a visible gap exists between the problems of synchronization in networks of passive heterogeneous agents, provided by a very simple algorithm,
and synchronization among general heterogeneous agents, which requires sophisticated model-based controllers. In this paper, we make a step towards filling this gap and show
that the conventional synchronization algorithm for passive agents~\citep{ChopraSpongBook} is applicable also to \emph{input-feedforward} passive (IFP)~\citep{Khalil,TorresHespanha:15} agents, provided that the couplings among them are sufficiently \emph{weak}. The class of IFP systems is much broader than the class of passive systems (and contains, in particular, all asymptotically stable linear systems). We demonstrate applications of our results to the design of cooperative adaptive cruise control (CACC) for platoons of automated vehicles and stability of a microscopic traffic flow model with delayed drivers' responses, both of which can be reduced to synchronization of IFP agents.

\section{Preliminaries}

In this section, we introduce basic concepts from graph theory and define input-feedforward passivity (IFP).

\subsection{Graphs and their connectivity properties}

A (weighted directed) graph is a triple $\g=(\V,\E,A)$, where $\V=\{v_1,\ldots,v_N\}$ stands for the set of \emph{nodes}, $\E\subset \V\times \V$ is a set of \emph{arcs} and $A=(a_{jk})_{j,k=1}^N$ is a non-negative \emph{adjacency matrix}, such that $a_{jk}>0$ if $(v_k,v_j)\in E$ and otherwise $a_{jk}=0$. We always assume that the number of nodes $N$ and their indices are fixed, so $V=\{1,\ldots,N\}$, there is a one-to-one correspondence between such graphs and their adjacency matrices $A\mapsto G[A]\dfb(V,E[A],A)$, where $E[A]\dfb\{(j,k):a_{kj}\ne 0\}$.
Henceforth all graphs have no self-loops $a_{jj}=0\,\forall j$. A graph is called \emph{undirected} if $A=A^{\top}$.
For any node $j$ we introduce the weighted \emph{in-} and \emph{out-degrees} $d_j^+[A]\dfb\sum_{k=1}^N a_{jk}$ and $d_j^-[A]\dfb\sum_{k=1}^N a_{kj}$.

A \emph{walk} connecting nodes $v$ and $v'$ is a sequence of nodes $v_{i_0}\dfb v,v_{i_1},\ldots,v_{i_{s-1}},v_{i_{s}}\dfb v'$ ($n\ge 1$) such that $(v_{i_{k-1}},v_{i_{k}})\in E$ for $k=1,\ldots,s$.
A graph is \emph{strongly connected} if a walk between any two distinct nodes exists. A graph is \emph{quasi-strongly connected}, or has a \emph{directed spanning tree}, if one of its nodes
is connected by walks to all other nodes. For an undirected graph these conditions are equivalent (such a graph is simply called \emph{connected}).

\subsection{Passivity and input-feedforward passivity}

Consider the dynamical system
\be\label{eq.nlin0}
\dot x(t)=f(x(t),u(t)),\,y(t)=h(x(t),u(t)),\quad t\ge 0,
\ee
where $x(t)\in\r^n$, $u(t)\in\r^m$ and $y(t)\in\r^m$ stand, respectively, for the state, control and output.

The system~\eqref{eq.nlin0} is \emph{passive}~\citep{Khalil,Willems1972common} if there exists a \emph{storage function} $V(x)\ge 0$ such that
\be\label{eq.pass1}
V(x(T))-V(x(0))\le \int_0^Ty(t)^{\top}u(t)\,ds\; \forall T\ge 0
\ee
(here $T$ varies in the interval where the solution exists). Assuming $V$ to be $C^1$-smooth, \eqref{eq.pass1} can be rewritten as
\be\label{eq.pass}
\dot V(x,u)=\frac{\partial V}{\partial x}f(x,u)\le h(x,u)^{\top}u\;\forall x\in\r^n,u\in\r^m.
\ee

In this paper, we primarily deal with systems, satisfying a ``relaxed'' passivity condition, defined as follows.
\begin{defn}\label{def.ifp}
The system~\eqref{eq.nlin0} is IFP($\alpha$) (input-feedforward passive with the passivity index $\alpha$) if it is passive with respect to the output $\tilde y=y+\alpha u$, i.e.
\be\label{eq.pass-f}
V(x(T))-V(x(0))\le \int_0^T\left(y(t)^{\top}u(t)+\alpha|u(t)|^2\right)\,dt.
\ee
\end{defn}
In the case $\alpha=0$ an IFP($\alpha$) system is passive; if $\alpha<0$ the condition~\eqref{eq.pass-f} is referred to as the \emph{strict} input passivity.
In this paper, we are primarily interested in systems that are not passive but IFP($\alpha$) with $\alpha>0$. Examples of such systems are discussed in Section~\ref{subsec.exam}.

Although this is not required by the formal definition, the conditions of passivity and IFP usually hold for systems with zero equilibrium: $f(0,0)=0$ and $h(0,0)=0$.
For systems without equilibria points a modification of passivity condition exists, referred to as the \emph{incremental passivity}~\citep{DePersisBayu:14,LiuHillZhao:15}.
Similarly, we introduce the incremental IFP condition.
\begin{defn}\label{def.iifp}
A dynamical system is said to be iIFP($\alpha$) (incrementally IFP($\alpha$)) if for any two solutions
$(x_1,u_1,y_1)$ and $(x_2,u_2,y_2)$ the respective deviations $\delta x=x_2-x_1$, $\delta u=u_2-u_1$, $\delta y=y_2-y_1$ satisfy the inequality
\be\label{eq.pass-f1}
V(\delta x(T))-V(\delta x(0))\le \int_0^T\left(\delta y^{\top}\delta u+\alpha|\delta u|^2\right)\,dt,
\ee
where $T$ belongs to the interval where both solutions exist. The function $V$ is called the incremental storage function.
\end{defn}

Obviously, for linear systems IFP($\alpha$)$\Longleftrightarrow$iIFP($\alpha$).

\section{Main Results: Synchronization Protocols for IFP agents}\label{sec.main}

Consider a group of $N$ agents obeying the equations:
\be\label{eq.nlin}
\dot x_j(t)=f_j(x_j(t),u_j(t)),\;y_j(t)=h_j(x_j(t)),\quad t\ge 0,
\ee
for $j\in\{1,\ldots,N\}$. Here $x_j(t)\in\r^{n_j}$, $u_j(t)\in\r^m$, $y_j(t)\in\r^m$ stand respectively for the $j$th agent's state, control and output.

In this paper, we study distributed protocols, synchronizing the outputs $y_j$ asymptotically or in $L_2$-norm.
\begin{defn}
Solutions $\{(x_j(t),u_j(t),y_j(t))\}_{j=1}^N$ of the systems~\eqref{eq.nlin}, defined on $t\in[0;\infty)$, are \emph{output synchronized} if
\be\label{eq.sync}
|y_i(t)-y_j(t)|\xrightarrow[t\to\infty]{} 0\quad\forall i,j=1,\ldots,N.
\ee
More specifically, the solutions are output synchronized with a predefined \emph{reference signal} $\bar y:[0;\infty)\to\r^m$ if
\be\label{eq.sync1}
|y_i(t)-\bar y(t)|\xrightarrow[t\to\infty]{} 0\quad\forall i=1,\ldots,N.
\ee
\end{defn}
\begin{defn}
Solutions $\{(x_j(t),u_j(t),y_j(t))\}_{j=1}^N$ of the systems~\eqref{eq.nlin}, defined for $t\ge 0$, are \emph{output $L_2$-synchronized} if
\be\label{eq.sync+}
\int_0^{\infty}|y_i(t)-y_j(t)|^2dt<\infty\quad\forall i,j=1,\ldots,N.
\ee
The solutions are output $L_2$-synchronized with a predefined reference signal $\bar y:[0;\infty)\to\r^m$ if
\be\label{eq.sync1+}
\int_0^{\infty}|y_i(t)-\bar y(t)|^2dt<\infty\quad\forall i=1,\ldots,N.
\ee
\end{defn}

In practice, the difference between the asymptotical and $L_2$-synchronization is minor.
Mathematically, none of these conditions implies the other one. However, in some special situations it is possible to prove that $L_2$-synchronization  implies asymptotical synchronization.
\begin{prop}\label{prop.barba}
Let $y_j(t)$ be absolutely continuous and $(\dot y_i-\dot y_j)\in L_p[0;\infty]$ for some $p>1$ and for any $i,j$. Then~\eqref{eq.sync+} implies~\eqref{eq.sync}. If, additionally, $\bar y(t)$ is absolutely continuous and $(\dot y_i-\dot{\bar{y}})\in L_p[0;\infty]\,\forall i$ then~\eqref{eq.sync1+} entails~\eqref{eq.sync1}.
\end{prop}

Proposition~\eqref{prop.barba}, as well as all other statements of this paper, is proved in Appendix.
In the following subsections we examine synchronization algorithms.

\subsection{Synchronization without reference signal}

We start examinating the linear controller:
\be\label{eq.proto-lin}
u_j(t)=\sum_{k=1}^Na_{jk}(y_k(t)-y_j(t)),
\ee
where $a_{jk}\ge 0$ are the \emph{coupling gains}. The matrix $A=(a_{jk})$ determines the interaction graph (or the network's \emph{topology}) $\g[A]$, where node $k$ is connected to node $j$ by an arc if and only if $a_{jk}\ne 0$, that is, the control input of agent $j$ is \emph{directly} influenced by the output of agent $k$.

It is widely known~\citep{Murray:07,RenBeardBook,Muenz:11} that single integrators $\dot y_j=u_j$, coupled via the protocol~\eqref{eq.proto-lin}
reach \emph{consensus} (that is, a common limit $y_*=\lim_{t\to\infty}y_j(t)$ exists) whenever $\g[A]$ has a directed spanning tree. Output synchronization~\eqref{eq.sync} is retained replacing single integrators by general \emph{passive} systems~\eqref{eq.nlin} and assuming strong connectivity of $\g[A]$~\citep[Theorem 8.3]{ChopraSpongBook}.
Our first result extends this to IFP agents.
\begin{thm}\label{thm.1}
Assume that agent $j$ (for $j=1,\ldots,N$) is IFP($\alpha_j$) with a storage function $V_j(x_j)\ge 0$.
Let $\g[A]$ be strongly connected and the couplings be ``weak'', i.e.
\be\label{eq.degree1}
\alpha_jd_j^+[A]=\alpha_j\sum_{k=1}^Na_{jk}<1/2\quad\forall j=1,\ldots,N.
\ee
Then the following statements hold.
\begin{enumerate}
\item Any solution of the system~\eqref{eq.nlin},\eqref{eq.proto-lin}, which is prolongable to $\infty$, is output $L_2$-synchronized~\eqref{eq.sync+};
\item Suppose that for any $j$ the function $V_j$ is \emph{radially unbounded} $\lim_{|x_j|\to\infty}V_j(x_j)=\infty$, the map $f_j$ is continuous and $h_j$ is $C^1$-smooth.
Then, any solution of the closed-loop system~\eqref{eq.nlin},\eqref{eq.proto-lin} is prolongable to $\infty$, bounded, and output synchronized~\eqref{eq.sync}.
\end{enumerate}
\end{thm}

The proofs of Theorem~\ref{thm.1} and other results of this section are given in Section~\ref{sec.proof}.
Note that in the case of $\alpha_j=0$ the inequalities~\eqref{eq.degree1} hold for any matrix $A$, and Theorem~\ref{thm.1} coincides with Theorem 8.3 in~\citep{ChopraSpongBook}.
We proceed with two remarks, regarding the assumptions.
\begin{rem}\label{rem.weak}
Unlike passive agents, for general IFP agents the requirement of weak coupling~\eqref{eq.degree1} cannot be disregarded, as demonstrated by the following example. For any $p,q>0$ the system:
\be\label{eq.agent2}
\dddot y_j(t)+p\ddot y_j(t)+q\dot y_j(t)=u_j(t)\in\r,\quad t\ge 0,
\ee
is IFP($\alpha$) with some $\alpha=\alpha(p,q)>0$ (c.f. Subsect.~\ref{subsec.exam}). Applying the protocol~\eqref{eq.proto-lin} with all-to-all coupling $a_{ij}=\varkappa>0\,\forall i,j$ to a group of identical agents~\eqref{eq.agent2}, output synchronization is guaranteed~\citep{Murray:07,LiDuanChen:10} only when the polynomial
$s^3+ps^2+qs+\varkappa(N-1)=0$
is Hurwitz. Accordingly to the Routh-Hurwitz criterion, this is possible only if $\varkappa(N-1)<pq$, i.e. the gain $\varkappa$ is small.
\end{rem}
\begin{rem}
Dealing with general heterogeneous agents, the condition of strong connectivity \emph{cannot} be replaced by the existence of a directed spanning tree in $\g[A]$. Consider, for instance, a pair ($N=2$) of harmonic oscillators
$$
\ddot \xi_1+\om_1^2\xi_1=u_1,\quad \ddot \xi_2+\om_2^2\xi_2=u_2,\quad\om_1\ne\om_2.
$$
that are passive with respect to the outputs $y_1=\dot\xi_1$ and $y_2=\dot\xi_2$. Consider the protocol $u_1=k(\dot \xi_2-\dot\xi_1), u_2=0$, which corresponds to the graph with $N=2$ nodes and the only arc $2\mapsto 1$. It can be shown that the system has a family of solutions $\xi_1(t)=\re[W(\imath\om_2)ce^{\imath\om_2t}]$, $\xi_2=\re[ce^{\imath\om_2t}]$, where $c\in\mathbb{C}$ is constant and $W(s)=ks/(s^2+ks+\om_1^2)$. The corresponding outputs are $y_1(t)=\re[\imath\om_2W(\imath\om_2)ce^{\imath\om_2t}]$ and $y_2(t)=\re[\imath\om_2ce^{\imath\om_2t}]$.
Since
\[
|W(\imath\om_2)|=\left|\frac{k\imath\om_2}{(\om_1^2-\om_2^2)+k\imath\om_2}\right|<1,
\]
the outputs are harmonic signals with the same frequency $\om_2$ but different amplitudes and cannot be synchronous.

Under some additional assumptions, synchronization over a quasi-strongly connected graph may be established by using the internal model control~\citep{IsidoriMarconi:2014}.
\end{rem}

\subsection{Reference-tracking synchronization}
We now consider the more complex problem of output synchronization with a reference signal~\eqref{eq.sync1}. In this paper we confine ourselves to a special situation:
when the desired trajectory is generated as the output of an agent for some appropriate control input and initial condition.
\begin{assum}\label{ass.out}
For any $j$ system~\eqref{eq.nlin} has a solution $(\bar x_j(t),\bar u_j(t),\bar y_j(t))$ such that $\bar y_j(t)\equiv \bar y(t)\forall t\ge 0$ (in particular, the solution is prolongable to $\infty$). At any time agent $j$ is aware of the value\footnote{If $\bar u_j(\cdot)$ is not unique, agent $j$ knows one of such solutions.} $\bar u_j(t)$, however the reference $\bar y(t)$ may be available only to a few ``dedicated'' agents.
\end{assum}

Assumption~\ref{ass.out} is often adopted implicitly or explicitly in reference-tracking synchronization problems. For \emph{linear} agents~\citep{LiDuanChen:10,LiuDePersisCao:15}
the reference signal $\bar y(t)$ is usually supposed to be an output of a reference system, whose model is known and included by the models of other agents.
Dealing with first-order integrator agents $\dot y_j=u_j$, Assumption~\ref{ass.out} implies that the agents know the derivative $\dot{\bar{y}}(t)$; this holds e.g. if $\bar y(t)=t\bar v+\bar y(0)$, where $\bar v$ is known, but the initial condition $\bar y(0)$ is uncertain.
A practical example of this type is discussed in Section~\ref{sec.cacc}.
Note that the solution $(\bar x_j(t),\bar u_j(t),\bar y_j(t))$ is not assumed to be asymptotically stable, so the control $u_j(t)=\bar u_j(t)$ \emph{does not} guarantee the reference signal tracking~\eqref{eq.sync1}. In general, only \emph{some of the agents} are able to measure the tracking error $\bar y(t)-y_j(t)$, whereas the remaining agents measure only deviation between theirs and their neighbors' outputs.

Consider the following modification of the algorithm~\eqref{eq.proto-lin}
\be\label{eq.proto-lin+}
u_i(t)=\bar u_i(t)+b_i(\bar y(t)-\bar y_i(t))+\sum_{j=1}^Na_{ij}(y_j(t)-y_i(t)).
\ee
Here $b_i>0$ if agent $i$ has access to the reference signal, and otherwise $b_i=0$. The following result is a counterpart of Theorem~\ref{thm.1} for reference-tracking synchronization.
\begin{thm}\label{thm.1+}
Let Assumption~\ref{ass.out} hold and further assume that: for all $j\in\{1,\ldots,N\}$ agent $j$ is iIFP($\alpha_j$), $\g[A]$ is strongly connected, at least one agent has access to the reference signal, i.e. $\sum_ib_i>0$, and the couplings are sufficiently weak, i.e.
\be\label{eq.degree1+}
\alpha_j(d_j^+[A]+2b_j)<1/2\quad\forall j=1,\ldots,N.
\ee
Then, the following two statements hold:
\begin{enumerate}
\item Any solution of the system~\eqref{eq.nlin},\eqref{eq.proto-lin+}, prolongable to $\infty$, is output $L_2$-synchronized with the reference signal~\eqref{eq.sync1+}; in particular,
$\int_0^{\infty}|\bar u(t)-u_j(t)|^2dt<\infty$.
\item If for all $j$ the functions $V_j$ are radially unbounded, the maps $f_j$ are $C^1$-smooth, the Jacobians $\frac{\partial f_j}{\partial x_j},\frac{\partial f_j}{\partial u_j}$ are uniformly bounded, and the maps $h_j$ are linear: $h_j(\xi_1-\xi_2)=h_j(\xi_1)-h_j(\xi_2)$;
 then, any solution of the closed-loop system~\eqref{eq.nlin},\eqref{eq.proto-lin+} is prolongable to $\infty$ and output synchronized~\eqref{eq.sync1} with the reference signal.
\end{enumerate}
\end{thm}

\subsection{Examples of IFP agents}\label{subsec.exam}

In this subsection, examples of IFP agents are provided.

\textbf{SISO agents with a pole at zero}

Consider a SISO system
\be\label{eq.SISO-pole0}
\begin{gathered}
s\rho(s)\zeta(t)=u(t)\in\r,\quad s\dfb\frac{d}{dt},\,\rho(\la)=\sum_{k=0}^r\rho_k\la^k;\\
y(t)=\eta(s)\zeta(t),\quad \eta(\la)=\sum_{k=0}^r\eta_k\la^k
\end{gathered}
\ee
\begin{lem}\label{lem.linear-pass}
Assume that $\rho(s)$ is a Hurwitz polynomial and $\eta_0\rho_0\ge 0$.
Then the system~\eqref{eq.SISO-pole0} is IFP($\alpha$) for sufficiently large $\alpha\ge 0$.
Denoting the transfer function from $u$ to $y$ by $W(\la)=\eta(\la)/(\la\rho(\la))$, the passivity index can be found as
\be\label{eq.alpfa-siso}
\alpha=-\inf_{\om\in\r}\re W(\imath\om).
\ee
\end{lem}

For instance, Lemma~\ref{lem.linear-pass} implies that the system~\eqref{eq.agent2} is IFP (in this case,
$\rho(\la)=\la^2+p\la+q$ is Hurwitz since $p,q>0$ and $y(t)=\xi(t)$).

\textbf{First-order delayed integrators}

Consider now a delayed system:
\be\label{eq.agent-delay0}
\dot y(t)=u(t-\alpha)\in\r^m.
\ee
Here $\alpha\ge 0$ is a constant delay and we assume, by definition, that $u(t)\equiv u_0(t)$ for $t\in [-\alpha;0]$,
where $u_0\in L_2([-\alpha;0]\to\r^m)$ is a given function. The vector $y(0)$ and the function $u_0$ are the initial conditions for the system~\eqref{eq.agent-delay0}.
Formally, our definition of IFP deals with
ordinary differential equations~\eqref{eq.nlin0} only and is not applicable to delay systems.
However, the following weaker condition holds for~\eqref{eq.agent-delay0}, see~\citep[p.141, proof of Lemma~7.3]{Pro14DelayBook}.
\begin{lem}\label{lem.delay-pass}
For any solution of~\eqref{eq.agent-delay0} one has
\be\label{eq.ifp-delay}
\int_0^{T}(y(t)^{\top}u(t)+\alpha|u(t)|^2)dt\ge -\mathcal{V}\quad\forall T\ge 0,
\ee
where $\mathcal{V}=\mathcal{V}(y(0),u_0(\cdot))\ge 0$ is independent of $T$.
\end{lem}

Lemma~\ref{lem.delay-pass} allows to extend the synchronization criteria to ensembles of agents~\eqref{eq.agent-delay}.
\begin{thm}\label{thm.delay-agent}
For a group of linear delayed agents
\be\label{eq.agent-delay}
\dot y_i(t)=u_i(t-\alpha_i),\quad i=1,\ldots,N,
\ee
the protocol~\eqref{eq.proto-lin} provides output synchronization~\eqref{eq.sync} and $L_2$-synchronization~\eqref{eq.sync+}, whenever
the graph $\g[A]$ is strongly connected and~\eqref{eq.degree1} holds.
\end{thm}
\begin{rem}
In the monograph~\cite{TianBook} a more general result is formulated without a complete proof (Theorem 7.10), 
stating that under assumptions of Theorem~\ref{thm.delay-agent} synchronization is retained if the graph is not strongly connected but has a directed spanning tree.
\end{rem}

\begin{rem}
Theorem~\ref{thm.1+} also holds for agents~\eqref{eq.agent-delay}. However, Assumption~\ref{ass.out} becomes impractical since each agent has to be aware of $\bar u_i(t)=\dot{\bar y}(t+\alpha_j)$ at time $t$, which makes the controller~\eqref{eq.proto-lin+} non-causal.
The protocol~\eqref{eq.proto-lin+} may still  be used in the case where the reference signal is linear $\bar y(t)=v_0t+\bar y(0)$ and $v_0$ is known, but $\bar y(0)$ is uncertain.
\end{rem}

\section{Synchronization in Vehicle Platooning and Traffic Flow Modeling}\label{sec.cacc}

In this section, we consider two practical applications of the synchronization criteria from Section~\ref{sec.main}.

\subsection{Stability of a microscopic traffic flow model}

A basic problem in vehicular traffic is the prevention of congestions and accidents.
\emph{Microscopic} traffic flow models are often employed
to represent the traffic flow as a result of cooperation between individual drivers.
Since the pioneering work of~\cite{Chandler}, the \emph{delay} in drivers reaction has been recognized as a crucial factor participating into the overall flow dynamics.
The simplest model of this kind \citep{Chandler,Niculescu:07} deals with $N$ vehicles, indexed $1$ through $N$, traveling along a common straight or
circular single lane road (their order remains unchanged since overtaking is not possible). Each driver is aiming to equalize his velocity of his own vehicle with that of its predecessor:
\be\label{eq.chandler}
\dot v_i(t) = u_i(t-\alpha),\,u_i(t)=K(v_{i-1}(t)-v_i(t)).
\ee
Here $v_i(t)$ is the speed of the $i$-th vehicle, $\alpha$ is the delay in its driver's action, and $K$ stands for the driver's ``sensitivity" to alterations of the relative velocity
of the predecessor vehicle. In the case of straight road, $v_0(t)\equiv v_0$ is the desired velocity with respect to the leading vehicle $1$; for a circular road, $v_0(t)\equiv v_N(t)$, i.e. vehicle $1$
follows vehicle $N$. A key issue addressed via this model \cite{Chandler} is that of the stability of the ``synchronous'' manifold: $v_1=\ldots=v_N$.

For the straight road case a necessary and sufficient condition for such a synchronization: $2\alpha K<1$ was found in~\eqref{eq.chandler}.
We extend this classical result to the traffic flow model with a general directed interaction topology and \emph{heterogeneous} delays and sensitivities of the drivers.
\be\label{eq.chandler+}
\dot v_i(t) = u_i(t-\alpha_i),\,u_i(t)=\sum_{j=1}^Na_{ij}(v_{j}(t)-v_i(t)),\;\forall i.
\ee
The model~\eqref{eq.chandler+} allows, in particular some drivers to respond to the change not only in the predecessor's, but also in the follower's velocity, or use the information about several predecessors and followers. The following theorem gives a criterion of velocity synchronization in~\eqref{eq.chandler+} under the assumption of a strongly connected topology, which holds e.g. for uni- and bidirectional ring coupling (circular road). The gain $a_{ij}\ge 0$ in~\eqref{eq.chandler+} stands for the sensitivity of driver $i$ to changes in the speed of vehicle $j$.
Theorem~\ref{thm.delay-agent}, applied to $y_i=v_i$, yields in the following corollary:
\begin{cor}
Suppose that the graph $\g[A]$ is strongly connected and~\eqref{eq.degree1} holds. Then the vehicles' velocities are asymptotically synchronized $v_i(t)-v_j(t)\xrightarrow[t\to\infty]{}0$.
\end{cor}

\subsection{An application to cooperative adaptive cruise control}

\begin{figure}[tp]
\includegraphics[width=\columnwidth]{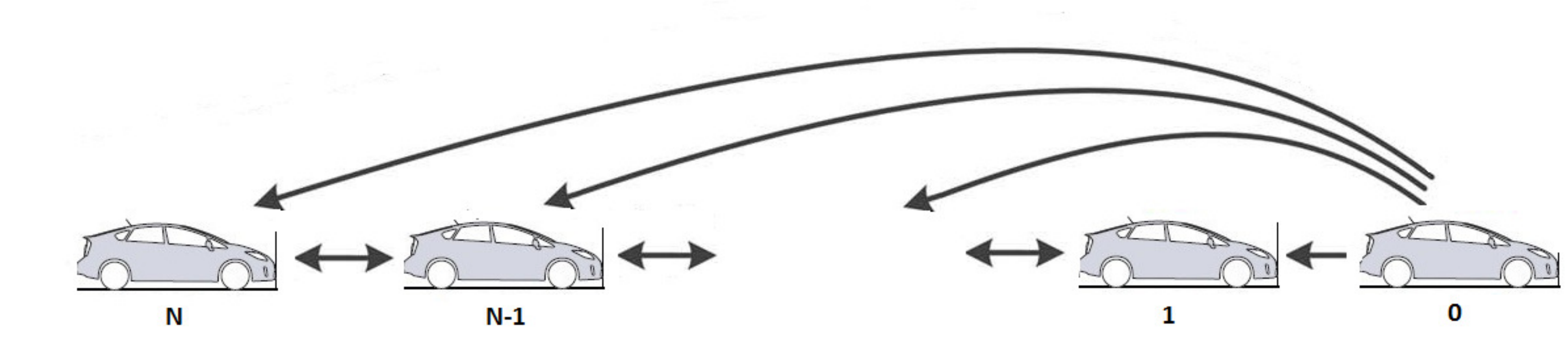}
\caption{Platoon of vehicles with bidirectional coupling.}\label{fig.topology}
\end{figure}

\begin{figure}[ht]
\center
\includegraphics[width=0.95\columnwidth]{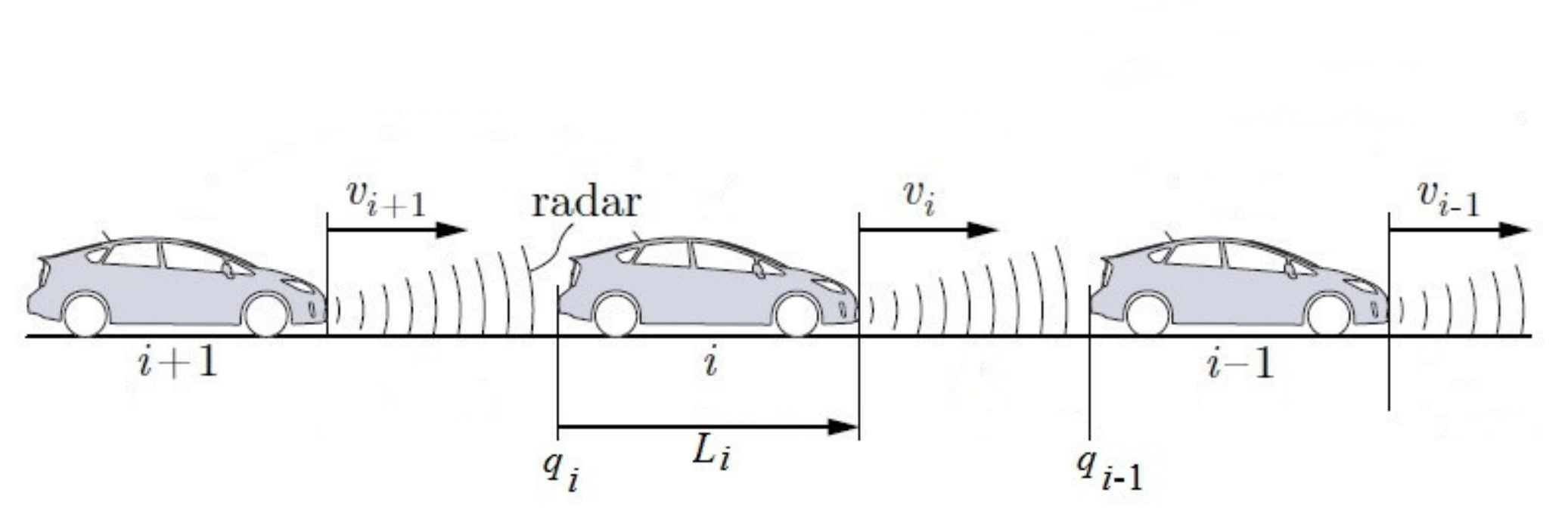}
\caption{Platoon of vehicles. Notation used in the text}\label{fig.platoon}
\end{figure}

In this subsection we demonstrate an application of Theorem~\ref{thm.1+} to the stability of a platoon of vehicles (Fig.~\ref{fig.platoon}), constituted by the leading vehicle $0$
and $N$ follower vehicles, indexed $1$ through $N$ (Fig.~\ref{fig.topology}).
Cooperative adaptive cruise control (CACC) system implements a control algorithm,
making each vehicle keep the safe distance to the predecessor and, provided that this safety constraint is satisfied, follow the leader's velocity. The interaction topology between the vehicles may be different~\citep{Zheng2016}; the most studied is a unidirectional topology, where each vehicle has information only about the predecessor.

In this subsection, we examine a CACC algorithm with \emph{bidirectional} interactions. The advantages of bidirectional platooning algorithms over unidirectional ones are discussed e.g. in~\citep{ZhangIoannou1999,Barooah2009,Zheng2016} (see also references therein); in many senses such algorithms are more robust against disturbances propagating through the platoon (``string-stable'').

We examine the CACC algorithm, proposed in~\citep{Barooah2009}. The leader's speed $v_0(t)\equiv v_0$ is broadcasted to every follower (Fig.~\ref{fig.topology}). Besides this, the vehicles $1$ through $N-1$ measure the distances to \emph{both} their predecessors and followers, and the rear vehicle $N$ measures the distance to its predecessor.
Denoting the position of vehicle $i$'s rear bumper by $q_i\in\r$ (see Fig.~\ref{fig.platoon}), the goal of the CACC algorithm is to keep the desired distance to the predecessor and the desired velocity, i.e.
\be\label{eq.cacc-goal}
q_{i-1}(t)-q_i(t)\xrightarrow[t\to\infty]{}s_i,\quad v_i(t)=\dot q_i(t)\xrightarrow[t\to\infty]{}v_0.
\ee

As usual in CACC problems~\citep{ZhangIoannou1999,Zheng2016}, the follower vehicles obey linear models
\be\label{eq.veh1}
\tau_i\dddot q_i+\ddot q_i=a_{i,des}(t),
\ee
where $a_{i,des}$ is the desired acceleration and $\tau_i$ is a time constant, depending on the vehicle's powertrain.
The vehicles $1$ through $N-1$ apply the following controller:
\be\label{eq.veh-proto1}
\begin{aligned}
a_{i,des}&(t)=\mu_i(v_0-v_i(t))+\eta_i(q_{i-1}(t)-q_i(t)-s_i)\\
&+\nu_i(q_{i+1}(t)-q_i(t)+s_{i+1}),\; 1\le i\le N-1,
\end{aligned}
\ee
Vehicle $N$ is controlled similarly, but has no follower
\be\label{eq.veh-proto2}
a_{N,des}(t)=\mu_N(v_0-v_N(t))+\eta_N(q_{N-1}(t)-q_N(t)-s_N).
\ee
\begin{thm}\label{thm.cacc}
Let $\mu_i\tau_i<\frac{1}{2}$ and $\eta_i,\nu_i>0$ satisfy
\be\label{eq.constants-vehi}
\frac{\mu_i^2}{2}>
\begin{cases}
\eta_i+\nu_i,\quad &1<i<N;\\
2\eta_1+\nu_1,\quad &i=1;\\
\eta_N,\quad &i=N.
\end{cases}\;\quad\forall i
\ee
Then the algorithm~\eqref{eq.veh-proto1},~\eqref{eq.veh-proto2} provides~\eqref{eq.cacc-goal}.
\end{thm}

The result of Theorem~\ref{thm.cacc} can be extended to some cases of nonlinear vehicles' dynamics, where
the inner-loop engine and torque controllers~\citep{ZhangIoannou1999} fail to attenuate the nonlinearities.
Notice that Theorem~\ref{thm.cacc} does not address the \emph{string stability} problem, i.e. the robustness of CACC against small disturbances in measurements as $N$ becomes large;
the analysis of string stability is based on other techniques and is beyond the scope of this paper.

\section{Conclusions and Future Work}

In this paper, we offer simple distributed protocols for synchronization of heterogeneous non-passive agents that
satisfy an IFP property. We apply the obtained results to analysis of microscopic traffic flow models and
CACC algorithms for heterogeneous platoons. The results can be extended to \emph{nonlinearly} coupled networks,
where the couplings satisfy the conditions of anti-symmetry and sector inequalities~\citep{ChopraSpongBook,Pro14DelayBook,ProMatv:15}, time-varying graphs
and antagonistic interactions among the agents~\citep{ProCao16-5}.
The results may be extended to discrete-time IFP agents.
The robustness of synchronization against measurements noises and communication delays are subjects of ongoing research.


\bibliography{consensus,cacc}

\appendix
\section{Technical Proofs}\label{sec.proof}

We start with several technical lemmas, used to prove Theorems~\ref{thm.1} and \ref{thm.1+}, and then proceed with proofs of Proposition~\ref{prop.barba}, Theorems~\ref{thm.1},~\ref{thm.1+},~\ref{thm.delay-agent} and~\ref{thm.cacc} and Lemma~\ref{lem.linear-pass}.

\subsection{Auxiliary lemmas}
The proofs of Theorems~\ref{thm.1},\ref{thm.1+} are based on the following lemma.
Consider two groups of vectors
$y_1,\ldots,y_N\in\r^m$ and $u_1,\ldots,u_N\in\r^m$, such that
\be\label{eq.tech}
u_i=\sum_{j=1}^Na_{ij}(y_j-y_i)\quad\forall i=1,\ldots,N.
\ee
\begin{lem}\label{lem.tech1}
Let the graph $\g[A]$ be strongly connected. Then the system of equalities~\eqref{eq.tech} entails that
\be\label{eq.tech2}
\sum_{i=1}^Np_iy_i^{\top}u_i=-\frac{1}{2}\sum_{i,j=1}^Np_ia_{ij}|y_j-y_i|^2,
\ee
where $(p_i)_{i=1}^N$ are constants, determined\footnote{In fact, $p^{\top}=(p_1,\ldots,p_N)$ stands for the non-negative left eigenvector of the weighted Laplacian matrix $L=L[A]$, corresponding to zero eigenvalue $p^{\top}L=0$. The strong connectivity of the graph implies that $p_i$ are strictly positive~\citep{ChopraSpongBook}.} by $A=(a_{ij})$.
\end{lem}
The proof of Lemma~\ref{lem.tech1} is a part of the proof of Theorem~8.5 in~\citep{ChopraSpongBook} and omitted here.

\begin{cor}\label{cor.tech}
Let the graph $\g[A]$ be strongly connected and the conditions~\eqref{eq.degree1} hold. Then~\eqref{eq.tech} entails that
\be\label{eq.tech2+}
\sum_{i=1}^Np_i\left(y_i^{\top}u_i+\alpha_i|u_i|^2\right)\le -\ve\sum_{i,j=1}^N|y_j-y_i|^2,
\ee
where $\ve>0$ is a constant, determined by the matrix $A$.
\end{cor}
\begin{pf}
Using the Cauchy-Schwartz inequality, one has
\[
\begin{aligned}
d_i^+[A]\alpha_i\sum_{j=1}^Na_{ij}|y_j-y_i|^2=\alpha_i\sum_{j=1}^Na_{ij}\sum_{j=1}^Na_{ij}|y_j-y_i|^2\ge\\
\ge \alpha_i\left|\sum_{j=1}^Na_{ij}^{1/2}a_{ij}^{1/2}(y_j-y_i)\right|^2\overset{\eqref{eq.tech}}{=}\alpha_i|u_i|^2\quad\forall i.
\end{aligned}
\]
By multiplying these inequalities by $p_i$, summing them up over $i$ and using~\eqref{eq.tech2}, it can be easily shown
that
\be\label{eq.tech3}
\sum_{i=1}^Np_i\left(y_i^{\top}u_i+\alpha_i|u_i|^2\right)\le -\sum_{i,j=1}^N\varkappa_i a_{ij}|y_j-y_i|^2,
\ee
where $\varkappa_i\dfb p_i(1/2-d_i^+[A]\alpha_i)\overset{\eqref{eq.degree1}}{>}0$.
The inequality~\eqref{eq.tech2+} with some $\ve=\ve(A)>0$ now follows from~\eqref{eq.tech3} since the matrix $(\varkappa_ia_{ij})_{i,j=1}^N$ corresponds to a strongly connected graph~\citep[Theorem~8]{Murray:04}.
\epf
\end{pf}

The proof of Theorem~\ref{thm.1+} also will use the following lemma.
\begin{lem}\label{lem.tech2}
Suppose that the system~\eqref{eq.nlin0} is IFP($\alpha$) with some $\alpha\ge 0$ and storage function $V$. Consider the constants $b\in (0;1/(2\alpha))$ and $\hat\alpha\dfb\alpha/(1-2\alpha b)$. Then the system~\eqref{eq.nlin0} is IFP($\hat\alpha$) for the same output $y$ and the new input $\hat u=u+by$ with the storage function
$\hat V(x)=V(x)/(1-2\alpha b)$. Moreover, for $\gamma\dfb b(1-\alpha b)/(1-2\alpha b)\ge 0$ one has
\be\label{eq.tech-ifp}
\hat V(x(T))-\hat V(x(0))\le \int_0^T\left(y^{\top}\hat u+\hat\alpha|\hat u|^2-\gamma|y|^2\right)dt
\ee
for any solution and $T\ge 0$. The statement retains its validity for iIFP system; in the latter case the vectors $x(t),y(t),u(t)$ in~\eqref{eq.tech-ifp} should be replaced
by the respective deviations $\delta x(t),\delta y(t),\delta u(t)$ between two solutions.
\end{lem}
\begin{pf}
The proof is straightforward from Definition~\ref{def.ifp}, substituting $u=\hat u-by$ into~\eqref{eq.pass-f} and noticing that
$$
y^{\top}u+\alpha|u|^2=(1-2\alpha b)\left(y^{\top}\hat u+\hat\alpha|\hat u|^2-\gamma|y|^2\right).
$$
The statement for iIFP system is proved in the same way, substituting $\delta u=\delta\hat u-b\delta y$ into~\eqref{eq.pass-f1}.\epf
\end{pf}

\subsection{Proof of Proposition~\ref{prop.barba}}

 Note first that an absolutely continuous function $\xi$, such that $\dot\xi\in L_p[0;\infty]$ with $p>1$, is uniformly continuous on $[0;\infty]$. This follows e.g. from the H\"older inequality, entailing that for $t\ge 0$ and $t'\ge t$
$$
|\xi(t)-\xi(t')|=\left|\int_t^{t'}\dot\xi(s)ds\right|\le (t'-t)^{q}\|\dot\xi\|_{L_p[t;t']},
$$
where $q=p/(p-1)$ (by definition, $q=1$ if $p=\infty$). If, additionally, $\xi\in L_2$, the Barbalat lemma~\citep{Khalil} implies that $\xi(t)\xrightarrow[t\to\infty]{}0$.
Proposition~\ref{prop.barba} now follows, applying this to, respectively, $\xi=y_j-y_i$ and $\xi=\bar y-y_i$.\epf

\subsection{Proof of Theorem~\ref{thm.1}.}

Let the conditions of Theorem~\ref{thm.1} hold. Introducing the stack vector $X(t)=[x_1(t)^{\top},\ldots,x_N(t)^{\top}]^{\top}$ and denoting
$V(X)\dfb\sum_{i=1}^Np_iV_i(x_i)$, the IFP($\alpha_i$) property of the agents~\eqref{eq.nlin} and Corollary~\ref{cor.tech} imply that
\be\label{eq.v-v}
\begin{aligned}
V(X(T))-V(X(0))\le\sum_{i=1}^N\int_0^Tp_i\left(y_i^{\top}u_i+\alpha_j|u_i|^2\right)dt\le\\
-\ve\sum_{i,j=1}^N\int_0^T|y_j(t)-y_i(t)|^2\,dt\le 0.
\end{aligned}
\ee
This implies statement (1): if the solution is defined for any $t\ge 0$, the solution is output $L_2$-synchronized.
To prove statement (2), notice that radial unboundedness of all storage functions $V_i(x_i)$ implies that $V(X)$ is also radially unbounded since $p_i>0\,\forall i$.
Since $V(x(T))\le V(x(0))$ for any $T$, the state vectors $x_i(t)$ are uniformly bounded, in particular, the solution does not escape to infinity in finite time and thus is infinitely prolongable. Recalling that the maps $h_i$ are continuous,
the outputs $y_i(t)$ are bounded; the same holds for $u_i(t)$ due to~\eqref{eq.proto-lin}. Since the maps $f_i$ are continuous, $\dot x_i(t)$ are bounded.
Recalling that $h_j$ is $C^1$-smooth, $\dot y_j(t)=h'(x_j(t))\dot x_j(t)$ is also bounded. Output synchronization~\eqref{eq.sync} now follows from Proposition~\ref{prop.barba}.\epf

\subsection{Proof of Theorem~\ref{thm.1+}}

Denoting $\tilde x_i(t)\dfb x_i(t)-\bar x_i(t)$, $\tilde u_i(t)\dfb u_i(t)-\bar u_i(t)$, $\tilde y_i(t)\dfb y_i(t)-\bar y_i(t)$, where $(\bar x_i(t),\bar u_i(t),\bar y_i(t))$ is the solution from Assumption~\ref{ass.out}. Recalling that $\bar y_i(t)\equiv \bar y(t)$, one has
\be\label{eq.tech4}
\tilde u_i(t)+b_i\tilde y_i(t)\overset{\eqref{eq.proto-lin+}}{=}\sum_{j=1}a_{ij}(\tilde y_j(t)-\tilde y_i(t)).
\ee
Denoting $\hat u_i\dfb\tilde u_i+b_i\tilde y_i$ and applying Lemma~\ref{lem.tech2} to the system~\eqref{eq.nlin}, $\alpha=\alpha_j$ and $b=b_j$, one arrives at
\be\label{eq.v-v+}
\hat V_i(\tilde x_i(T))-\hat V_i(\tilde x_i(0))\le \int_0^T\left(\tilde y^{\top}_i\hat u_i+\hat\alpha_i|\hat u_i|^2-\gamma_i|\tilde y_i|^2\right)dt,
\ee
where $\hat\alpha_i=\alpha_i/(1-2b_i\alpha_i)$ and $\gamma_i>0$ if and only if $b_i>0$.
The inequalities~\eqref{eq.degree1+} imply that $d_j^+[A]\hat\alpha_j<1/2$. Thus retracing the proof of Corollary~\ref{cor.tech}, \eqref{eq.tech4} implies that
\be\label{eq.tech2++}
\sum_{i=1}^Np_i\left(\tilde y_i^{\top}\hat u_i+\hat\alpha_i|\hat u_i|^2\right)\le -\ve\sum_{i,j=1}^N|y_j-y_i|^2,
\ee
where $p_i>0$ and $\ve>0$ depend on $A$, $\alpha_i$ and $b_i$. Introducing the stack vector $\tilde X(t)=[\tilde x_1(t)^{\top},\ldots,\tilde x_N(t)^{\top}]^{\top}$ and
the storage function $\hat V(\tilde X)=\sum_ip_i\hat V_i(\tilde x_i)$, we obtain
\be\label{eq.tech5}
\begin{split}
\hat V(\tilde X(T))-\hat V(\tilde X(0))\le -\ve\sum_{i,j=1}^N\int_0^T|\tilde y_j(t)-\tilde y_i(t)|^2\,dt-\\-\sum_ip_i\gamma_i\int_0^T|\tilde y_i(t)|^2dt=
-\int_0^T\mathcal F(\tilde y_1(t),\ldots,\tilde y_N(t))dt.
\end{split}
\ee
Here $\mathcal F(y_1,\ldots,y_N)\dfb \ve\sum_{i,j=1}^N|y_j-y_i|^2+\sum_ip_i\gamma_i|y_i|^2$ is a quadratic form; since $\gamma_i>0$ for at least one $i$, this form is \emph{positive definite}
and thus $\mathcal F(y_1,\ldots,y_N)\ge \ve_0(|y_1|^2+\ldots+|y_N|^2)$ for sufficiently small constant $\ve_0>0$.
The end of the proof retraces the proof of Theorem~\ref{thm.1}. If a solution exists,~\eqref{eq.tech5} implies that $\tilde y_i(t)=\bar y(t)-y_i(t)$ is $L_2$-summable and hence $\tilde u_i(t)=\bar u(t)-u_i(t)$ is $L_2$-summable, which proves statement (1). To prove statement (2), notice that  since $V_i$ are radially unbounded,~\eqref{eq.tech5} implies that the deviation of the solutions
$\tilde X(t)$ remains bounded; since $\bar x_i(t)$ is globally defined, $X(t)$ cannot grow unbounded in finite time. Thus the solution is prolongable to $\infty$.
Recalling that $h_i$ is a linear map, the function $\tilde y_i(t)=y_i(t)-\bar y(t)=h_i(x_i(t)-\bar x_i(t))=h_i(\tilde x_i(t))$ is uniformly bounded for any $i$.
Thus $\tilde u_i(t)$ is bounded due to~\eqref{eq.tech4} and
\[
\begin{aligned}
|\dot{\tilde{x}}_i(t)|&=|f_i(x_i(t),u_i(t))-f_i(\bar x_i(t),\bar u_i(t))|\le \\
&\le |\tilde{x_i}(t)|\sup\limits_{x_i,u_i}\left|\frac{\partial f_i(x_i,u_i)}{\partial u_i}\right|+|\tilde{u_i}(t)|\sup\limits_{x_i,u_i}\left|\frac{\partial f_i(x_i,u_i)}{\partial u_i}\right|
\end{aligned}
\]
(here suprema are taken over the space of all possible vectors $x_i\in\r^{n_i}$, $u_i\in\r^m$).
Thus the functions $\dot{\tilde{x}}_i(t)$ are also uniformly bounded;
using the linearity of $h_i$, the same holds for $\dot{\tilde{y}}_i(t)=h_i(\dot{\tilde{x}}_i(t))$. Applying Proposition~\ref{prop.barba}, the solutions are output synchronized~\eqref{eq.sync1}.

\subsection{Proof of Lemma~\ref{lem.linear-pass}}

Lemma~\ref{lem.linear-pass} is immediate from a more general result,
which in turn is implied by the standard positive real lemma.
Consider a linear SISO system
\be\label{eq.linear}
\dot x=Px+Qu,\quad y=Rx+Su
\ee
and let $W(\la)\dfb S+R(\la I-P)^{-1}Q$ stand for its scalar transfer function.

\begin{lem}\label{lem.prl}
Let the system~\eqref{eq.linear} be controllable and observable. Then it is IFP($\alpha$) for some $\alpha$ if and only if the following conditions hold
\begin{enumerate}
\item the matrix $P$ has no strict unstable eigenvalues: $\det(\la I-P)\ne 0$ when $\re\la>0$;
\item all imaginary eigenvalues (if they exist) are simple, at any such eigenvalue $\la=\imath\om_0$ the residual
is non-negative $\lim_{\la\to\imath\om_0}(\la-\imath\om_0)W(\la)\ge 0$;
\item $\re W(\imath\om)+\alpha\ge 0$ for any $\om\in\r$ such that $\det(\imath\om I-P)\ne 0$.
\end{enumerate}
\end{lem}
\begin{pf}
As was noticed in Section~2, the IFP($\alpha$) condition is equivalent to passivity of system~\eqref{eq.linear} with respect to the new input $\tilde y=y+\alpha u$. The statement of Lemma~\ref{lem.prl} is immediate, applying the result of~\citep[Theorem~1 in Part~2]{Willems1972common} (the positive real lemma) to the respective transfer function $\tilde W(\imath\om)=\alpha I_m+W(\imath\om)$.\epf
\end{pf}

Notice that if the condition (1) and (2) in Lemma~\ref{lem.prl} hold then (3) is valid for sufficiently large $\alpha>0$. Indeed, the function $\re W(\imath\om)=[W(\imath\om)+W(-\imath\om)]/2$ (where $\om\in\r$) is bounded as $\omega\to\infty$ and in the vicinity of any imaginary pole due to statement (2), thus this function is bounded and, in particular, semi-bounded from below.

Proof of Lemma~\ref{lem.linear-pass} is now obvious from Lemma~\ref{lem.prl} since the system~\eqref{eq.SISO-pole0} can be rewritten as a controllable and observable system~\eqref{eq.linear} with a single imaginary pole $\la=0$.
Such a system satisfies condition (1) in Lemma~\ref{lem.prl}, and (2) also holds since $\lim_{\la\to 0}\la W(\la)=\eta_0/\rho_0\ge 0$.

\subsection{Proof of Theorem~\ref{thm.delay-agent}}

Any solution of the linear time-invariant closed-loop system~\eqref{eq.agent-delay},~\eqref{eq.proto-lin} is infinitely prolongable.
A closer look at the proof of statement (1) in Theorem~\ref{thm.1} reveals that the IFP($\alpha_i$) condition can be replaced by a weaker condition
\ben
\int_0^{T}(y_i(t)^{\top}u_i(t)+\alpha_i|u_i(t)|^2)dt\ge -\mathcal{V}_i\quad\forall T\ge 0,
\een
where $\mathcal{V}_i\ge 0$ is some constant (for an IFP($\alpha_i$) agent~\eqref{eq.nlin} with storage function $V_i$, the 
latter inequality holds for $\mathcal{V}_i=V_i(x_i(0))$). Thanks to Lemma~\ref{lem.delay-pass}, statement (1) is valid for the agents~\eqref{eq.agent-delay} and thus any solution is
output $L_2$ synchronized. Since $u_i\in L_2[0;\infty]$ for any $i$, $\dot y_i\in L_2[0;\infty]$ due to~\eqref{eq.agent-delay} and thus outputs are also asymptotically synchronized due to Proposition~\ref{prop.barba}.\epf

\subsection{Proof of Theorem~\ref{thm.cacc}}

Introducing the control inputs $u_i\dfb a_{i,des}+\mu_i v_i$ and outputs $y_i=q_i+(s_{i}+s_{i-1}+\ldots+s_1)$, the closed-loop system~\eqref{eq.veh1},~\eqref{eq.veh-proto1},~\eqref{eq.veh-proto2} boils down to a group of
agents
\be\label{eq.agent-veh}
\tau_i\dddot y_i+\ddot y_i+\mu_i\dot y_i=u_i,\quad i=1,\ldots,N,
\ee
coupled through~\eqref{eq.proto-lin+}. Here $\bar u_i(t)=\mu_iv_0$, $\bar y(t)=q_0(t)$ and
\[
b_i=
\begin{cases}
\eta_1,\,i=1\\
0,\,i>1
\end{cases},\quad a_{ij}=\begin{cases}
\eta_i,\quad i>1,\,j=i-1\\
\nu_i,\quad i<N,\,j=i+1\\
0,\quad\text{otherwise}.
\end{cases}
\]
Using Lemma~\ref{lem.linear-pass}, the agent~\eqref{eq.agent-veh} is IFP($1/\mu_i^2$) since
\[
\re \frac{1}{\tau_i(\imath\om)^3+(\imath\om)^2+\mu_i(\imath\om)}=-\frac{1}{\mu_i^2+(1-2\tau_i\mu_i)\om^2+\om^4}.
\]
A straightforward computation shows that~\eqref{eq.constants-vehi} implies~\eqref{eq.degree1+}.
Obviously, the graph $\g[A]$ is a bidirectional chain and thus is strongly connected.
Hence, the outputs are $L_2$-synchronized~\eqref{eq.sync1+} and $u_i-\bar u_i\in L_2$. Denoting $\ve_i\dfb v_i-v_0$, one has $\dot y_i=v_i=\ve_i+v_0$, $\ddot y_i=\dot\ve_i$, $\dddot y_i=\ddot\ve_i$ and hence
$$
\tau_i\ddot\ve_i+\dot\ve_i+\mu_i\ve_i=u_i-\mu v_0=u_i-\bar u_i\in L_2[0;\infty],
$$
so that $\ve_i=\dot y_i-\dot{\bar{y}}\in L_2$ and $\dot\ve_i\in L_2$. Applying Proposition~\eqref{prop.barba}, one proves that~\eqref{eq.sync} holds and $\ve_i(t)\to 0$ as $t\to\infty$, which implies~\eqref{eq.cacc-goal}.\epf

\end{document}